# *Tract Orientation and Angular Dispersion Deviation Indicator* (*TOADDI*): A framework for single-subject analysis in diffusion tensor imaging

Cheng Guan Koay[1,3,4], Ping-Hong Yeh[1,2], John M. Ollinger[1], M. Okan İrfanoğlu[2,3], Carlo Pierpaoli[3], Peter J. Basser[3], Terrence R. Oakes[1], Gerard Riedy[1,5]

[1]National Intrepid Center of Excellence (NICoE), Bethesda, MD
[2]The Henry M. Jackson Foundation for the Advancement of Military Medicine, Bethesda, MD
[3]Section on Tissue Biophysics and Biomimetics, NICHD, National Institutes of Health, Bethesda, MD
[4]NorthTide Group, LLC
[5]National Capital Neuroimaging Consortium, Bethesda, MD

*Corresponding author:*
*Cheng Guan Koay, PhD*
*National Intrepid Center of Excellence*
*Walter Reed National Military Medical Center,*
*Bethesda, MD, USA*
*E-mail: guankoac@mail.nih.gov*
*Phone: 301-319-3767*

**ABSTRACT**

The purpose of this work is to develop a framework for single-subject analysis of diffusion tensor imaging (DTI) data. This framework is termed **T**ract **O**rientation and **A**ngular **D**ispersion **D**eviation **I**ndicator (TOADDI) because it is capable of testing whether an individual tract as represented by the major eigenvector of the diffusion tensor and its corresponding angular dispersion are significantly different from a group of tracts on a voxel-by-voxel basis. This work develops two complementary statistical tests based on the elliptical cone of uncertainty, which is a model of uncertainty or dispersion of the major eigenvector of the diffusion tensor. The orientation deviation test examines whether the major eigenvector from a single subject is within the average elliptical cone of uncertainty formed by a collection of elliptical cones of uncertainty. The shape deviation test is based on the two-tailed Wilcoxon-Mann-Whitney two-sample test between the normalized shape measures (area and circumference) of the elliptical cones of uncertainty of the single subject against a group of controls. The False Discovery Rate (FDR) and False Non-discovery Rate (FNR) were incorporated in the orientation deviation test. The shape deviation test uses FDR only. TOADDI was found to be numerically accurate and statistically effective. Clinical data from two Traumatic Brain Injury (TBI) patients and one non-TBI subject were tested against the data obtained from a group of 45 non-TBI controls to illustrate the application of the proposed framework in single-subject analysis. The frontal portion of the superior longitudinal fasciculus seemed to be implicated in both tests (orientation and shape) as significantly different from that of the control group. The TBI patients and the single non-TBI subject were well separated under the shape deviation test at the chosen FDR level of 0.0005. TOADDI is a simple but novel geometrically based statistical framework for analyzing DTI data. TOADDI may be found useful in single-subject, graph-theoretic and group analyses of DTI data or DTI-based tractography techniques.

## 1. INTRODUCTION

Diffusion Tensor Imaging (DTI) (Basser et al., 1994a, b; Pierpaoli et al., 1996) is an important and noninvasive magnetic resonance imaging (MRI) technique used in clinical and research studies of brain white matter architecture. Many aspects of DTI have been studied extensively, from the diffusion tensor model itself (Basser, 2002; Basser et al., 1994b; Basser and Pajevic, 2003; Koay and Özarslan, 2013; Stejskal, 1965) and its higher-order generalizations (Anderson, 2005; Descoteaux et al., 2007; Descoteaux et al., 2011; Jian et al., 2007; Liu et al., 2004; Özarslan and Mareci, 2003), to optimal experimental designs (Cook et al., 2007; Deriche et al., 2009; Dubois et al., 2006; Jones et al., 1999; Koay et al., 2011; Koay et al., 2012) and its inverse problem at various levels of complexity (Andersson, 2008; Basser et al., 1994a; Chang et al., 2005; Chang et al., 2012; Koay et al., 2006; Mangin et al., 2002; Maximov et al., 2011; Veraart et al., 2013; Wang et al., 2004). DTI continues to inspire new analyses, developments, refinements and extensions (Caruyer et al., 2013; Hutchinson et al., 2012; Koay, 2009; Koay, 2014; Koay et al., 2011; Koay et al., 2009a; Koay et al., 2012; Koay et al., 2009b; Wu et al., 2004). Uncertainty quantification (Anderson, 2001; Behrens et al., 2003; Beltrachini et al., 2013; Chang et al., 2007; Jeong and Anderson, 2008; Jones, 2003; Jones and Pierpaoli, 2005; Koay et al., 2007; Koay et al., 2008; Lazar and Alexander, 2003, 2005; Lazar et al., 2005; Poonawalla and Zhou, 2004) in DTI is another important area of research with wide-ranging implications to tractography (Barbieri et al., 2011; Basser et al., 2000; Conturo et al., 1999; Mori et al., 1999; Pajevic et al., 2002; Poupon et al., 2000; Poupon et al., 2001) and data analyses (longitudinal, single-subject, group or graph-theoretic, e.g., (Rubinov and Sporns, 2010)).

The purpose of this work is to develop a statistical and computational framework for testing whether an individual principal direction (i.e., major eigenvector of a diffusion tensor) and its corresponding elliptical cone of uncertainty (COU, see Methods for a brief review)(Koay et al., 2007; Koay et al., 2008) are different from a control group of diffusion-tensor-derived major eigenvectors and their corresponding elliptical COUs. Two key motivations behind the work are the pursuit of personalized or precision medicine for individual patients and to address the current inadequacy of group analysis of DTI data in clinical TBI studies.

TBI presents unique challenges to diagnosis due to the heterogeneity of injury source and manifestation across subjects, and mild TBI (mTBI) in particular is challenging due to the subtle imaging findings. Because of the varying degrees of severity of the injuries and the spatial heterogeneity of the affected regions not only in an individual TBI patient but also across TBI patients, there is a clear need to develop diagnostic methods that are effective on an individual basis. Since no two clinical TBI cases are alike, it may not be appropriate to group TBI data for voxel-wise analyses. The key concern is that the averaging effect of the group analysis may smooth out salient anomalies in a single patient and cause them to appear as normal variations within the *averaged* TBI data, i.e., it is very unlikely to build up a statistical group difference at a single voxel if there is rarely a lesion from more than one patient in a voxel.

This framework builds upon as well as extends the capability of our previously proposed uncertainty quantification framework for DTI (Koay et al., 2007; Koay et al., 2008). Our work on an analytical error propagation framework (Koay et al., 2007; Koay et al., 2008) for DTI is a culmination of prior works by others (Anderson, 2001; Jeong and Anderson, 2008; Jeong et al., 2005; Lazar and Alexander, 2003, 2005; Lazar et al., 2005) and our group (Koay and Basser, 2006; Koay et al., 2006). In brief, three independent studies adopted a perturbation-based error analysis (Anderson, 2001; Basser, 1997; Chang et al., 2007) to study uncertainty in fiber orientation and in tensor-derived quantities. These studies developed their respective error analyses from the linear model of the diffusion tensor. Our prior experience with signal and noise characterization in MRI (Koay and Basser, 2006) together with the observations made by Jones et al.(Jones and Basser, 2004) on the effects of noise on tensor-derived quantities led to the adoption of the nonlinear least squares model of the diffusion tensor as the model of choice (Koay et al., 2006) for error propagation (Koay et al., 2007). The most notable difference between the perturbation-based error analysis of Basser (Basser, 1997) or Chang et. al. (Chang et al., 2007) and our error propagation framework is that the former did not incorporate the elliptical COU into the formulation while such a feature is inherent in our framework (Koay et al., 2007). It is important to note that many studies (Basser, 1997; Behrens et al., 2003; Chang et al., 2007; Jones, 2003; Parker et al., 2003; Polders et al., 2011) have adopted the circular COU for modeling the uncertainty in the major eigenvector of the diffusion tensor even though converging and empirical evidence showed that the uncertainty of the major eigenvector is generally elliptical (Jeong et al., 2005; Lazar and Alexander, 2005; Lazar et al., 2005). The key reason for the lack of such an important feature (the elliptical COU) in DTI error analysis in the works of Basser and Chang is due to the absence of the covariance matrix of the major eigenvector in their formulations, which was recently demonstrated to be obtained from perturbation analysis through a simple reformulation; see (Koay et al., 2008) for the connection between our analytical error propagation framework and the reformulated perturbation-based error analysis.

The covariance matrix of the major eigenvector of the diffusion tensor provides the necessary information to construct the elliptical COU and related scalar measures such as the normalized areal and circumferential measures of the elliptical COU (Koay et al., 2008). While it is relatively easy to visualize or quantify a single elliptical COU within a voxel, it is nontrivial to test whether an individual major eigenvector or the boundary points of its elliptical cone is within the mean elliptical cone of uncertainty formed by a collection of elliptical COUs. This inclusion-exclusion test is a multivariate and geometric problem and is highly relevant to each phase of data analysis, from exploratory investigation to testing of hypotheses. While it is not hard to test the shape characteristics of an individual elliptical COU against those of a collection of elliptical COUs, the main obstacle is in the normalization of a family of elliptical COUs. This problem is solved through a novel utilization of the tensor normalization technique proposed by Zhang et. al. (Zhang et al., 2006). While there are several published approaches to performing voxel-based analysis for DTI (Schwarz et al., 2014; Smith et al., 2006; Zhang et al., 2006), it is not the intent of this work to compare relative merits of these approaches. Challenges of voxel-based analysis have been extensively studied by (Ashburner and Friston, 2000) and one of the key challenges is registration errors, which may cause unnecessary deformations or distortions that do not appear in the original image. One of the ways to minimize the registration errors in statistical inference is to through skeletonization as suggested in Tract-Based Spatial Statistics (TBSS) of (Smith et al., 2006) and our approach is to focus only on voxels within the white matter region that satisfy further criteria such as goodness of fit.

To the best of our knowledge, no proposed framework addresses the problems raised above. We propose a novel and conceptually simple statistical and computational pipeline to address these problems. Particular attention is paid to the antipodally symmetric nature of the elliptical cones in order to develop an unbiased orientation of the elliptical COUs. The inverse Gnomonic projection (Coxeter, 1989; Koay et al., 2008) of the elliptical COU onto the unit sphere is used to ensure the tests are well-defined.

The proposed framework, which we termed it as **T**ract **O**rientation and **A**ngular **D**ispersion **D**eviation **I**ndicator (TOADDI), is capable detecting a statistically and orientationally significant deviation of a principal diffusion direction as compared to those from a group of controls on a voxel-by-voxel basis. It can also detect any statistically significant deviation in the shape characteristics such as the normalized areal measure and the normalized circumferential measure of the elliptical cone as compared to those from the control group. While it would beyond the scope of this paper, which focuses mainly on the proposed methodological framework, to pursue a thorough investigation of all the available clinical TBI cases acquired at the National Intrepid Center of Excellence (NICoE), and in the interest of expediting the process of translational research in clinical TBI studies, we believe it would be beneficial to illustrate the application of the proposed framework on a few clinical TBI cases with complete details of the processing pipeline provided.

Here, we present two key contributions of the present work. First, this work incorporates the methods of False Discovery Rate (FDR) (Benjamini and Hochberg, 1995; Benjamini and Yekutieli, 2001; Yekutieli and Benjamini, 1999), popularized by Genovese et. al. (Genovese et al., 2002) within the context of neuroimaging to control the proportion of false positives in multiple comparison problems, and False Non-discovery Rate (FNR) (Genovese and Wasserman, 2002) within the proposed generalized multivariate test for testing orientation differences. Second, this work shows how differences in shape characteristics - such as area and circumference - of the elliptical cones between a single subject and a control group can be tested using the nonparametric test of Wilcoxon-Mann-Whitney (WMW) (Mann and Whitney, 1947; Wilcoxon, 1945).

Finally, we should mention that the proposed methodology is not a new tissue model at the neuronal level, e.g., (Zhang et al., 2012). While the angular dispersion of the major eigenvector may be affected by the underlying microstructure as well as noise, the geometrical characteristics of the angular dispersion may only be a gross representation of an admixture of the underlying biophysical processes. Further investigation is needed to find out whether the angular dispersion of the major eigenvector corresponds well with the underlying biophysical processes at the neuronal level. A comparison with some of the existing tissues models would be very interesting. This comparative study is being planned but it is beyond the scope of this work.

## 2. METHODS

*2.1 Brief Review of the Elliptical COU in DTI*

The elliptical COU quantifies the uncertainty of all three eigenvectors of the diffusion tensor. It is based on propagating errors from the noisy diffusion-weighted signals through the nonlinear least squares estimation model (Koay et al., 2006; Koay et al., 2007; Koay et al., 2008). The first order approximation of the uncertainty in any eigenvector of interest of the diffusion tensor is in a form of an ellipse around that eigenvector, see Figure 1A. Specifically, the ellipse on the u-v plane, which is the plane of Gnonomic projection, is orthogonal to the eigenvector of interest and the orientation of this eigenvector is pointing exactly at the center of the ellipse. Without loss of generality, the major eigenvector will be used throughout this work. We should mention that the framework is applicable to other eigenvectors as well.

The Gnomonic projection (Coxeter, 1989) maps a point, $\mathbf{s}$, on the unit sphere to a point, $\mathbf{p}$, on the u-v plane that is tangent to the sphere at z=1 on the z-axis, see Figure 1A. The inverse Gnomonic projection maps $\mathbf{p}$ to $\mathbf{s}$. Measures such as the area and the circumference of the ellipse are unbounded when the ellipse is lying on the u-v plane. However, these measures can be normalized by computing the equivalence of these measures on the surface of the unit sphere through the inverse Gnomonic projection as proposed in (Koay et al., 2008), see Figure 1B. Similarly, the visualization of the uncertainty of the eigenvector on the surface of the unit sphere as shown in Figure 1C is more convenient because such an approach avoids the issue of overlapping cones from the neighboring voxels as shown in Figure 2 of (Jones, 2003). Note that the elliptical cone is antipodally symmetric because the eigenvectors are axial vectors.

From the covariance matrix of the major eigenvector as determined through the analytical error propagation framework (see Appendix A), it can be shown (Koay et al., 2008) that the approximate $1-\alpha$ confidence region formed by the elliptical COU of the major eigenvector, $\mathbf{q}_1(\hat{\gamma})$, can be characterized by the major and minor axes of the ellipse in vectorial representation, which are given by $\sqrt{2F(2,n-7;\alpha)\omega_1}\,\mathbf{c}_1$ and $\sqrt{2F(2,n-7;\alpha)\omega_2}\,\mathbf{c}_2$, where $(\omega_1,\mathbf{c}_1)$ and $(\omega_2,\mathbf{c}_2)$ are respectively the major and medium eigenvalue-eigenvector pairs of the covariance matrix of the major eigenvector of the diffusion tensor, $F(2,n-7;\alpha)$ is the upper $\alpha$ quantile for the $F$ distribution with $2$ and $n-7$ degrees of freedom and $n$ is the number of diffusion-weighted measurements. Note that the minor eigenvalue-eigenvector pair of the covariance matrix of the major eigenvector is $(0,\mathbf{q}_1(\hat{\gamma}))$ and $\hat{\gamma}$ is an estimated parameter vector composed of the natural logarithm of the non-diffusion weighted signal and the vectorized diffusion tensor elements as defined in Appendix A and (Koay et al., 2006). Formulating the uncertainty in the major eigenvector in a form of an elliptical cone is ideal for visualization and computation of scalar measures such as the areal and circumferential measures. However, this formulation is cumbersome for the analysis to be proposed this work.

Even though the above formulation is in fact derived from a multivariate test (Koay et al., 2008), which is in a form of an inequality

$$(\mathbf{q}_1-\mathbf{q}_1(\hat{\gamma}))^T\mathbf{\Sigma}_{\mathbf{q}_1}^{+}(\hat{\gamma})(\mathbf{q}_1-\mathbf{q}_1(\hat{\gamma}))\leq 2F(2,n-7;\alpha)\,,\qquad\qquad\text{[1]}$$

where $\mathbf{\Sigma}_{\mathbf{q}_1}^{+}(\hat{\gamma})$ is the pseudoinverse of the covariance matrix of the major eigenvector and the covariance matrix of the major eigenvector, $\mathbf{\Sigma}_{\mathbf{q}_1}(\hat{\gamma})$, is derived from the error propagation framework (Koay et al., 2007; Koay et al., 2008), the importance of this multivariate test was not fully emphasized in our earlier work (Koay et al., 2008). For completeness, Appendix A contains a step-by-step guide on the computation of the covariance matrix of the major eigenvector.

*2.2 Novel FDR-FNR-based Orientation Deviation Indicator*

Given a collection of $N$ elliptical COUs, which in turn depends a collection of covariance matrices of the major eigenvector, $\left\{ \Sigma_{\mathbf{q}_1}(\hat{\gamma}_i) \right\}_{i=1}^{N}$, an orientation deviation test can be constructed from Eq.[1] by replacing the pseudoinverse of the covariance matrix of the major eigenvector with the pseudoinverse of the mean covariance matrix of the major eigenvector and $\mathbf{q}_1(\hat{\gamma})$ by a mean major eigenvector. This is our proposed generalized multivariate test for orientation differences.

In general, there are several approaches for averaging covariance matrices such as the geometric mean approach based on the Riemannian distance of symmetric positive definite (SPD) matrices proposed by Moakher (Moakher, 2006), the arithmetic mean approach or the mean dyadics approach (Koay et al., 2007). Unfortunately, the 3x3 covariance matrix of the major eigenvector as generated from the propagation of errors is rank deficient, i.e., its matrix rank is two rather than three hence the matrix is not SPD. This deficiency excludes the use of the geometric mean approach of Moakher because this approach requires the covariance matrices to be SPD. Thus, we are left with the arithmetic mean approach and the mean dyadics approach.

Preliminary analysis by simulation, see the Results section, shows that the arithmetic mean approach performs better than the mean dyadics approach in terms of the relative errors in estimating the areal and circumferential measures of the elliptical COU. Based on this preliminary analysis, we adopted the arithmetic mean approach throughout this work.

The arithmetic mean approach is simply the matrix sum of the covariance matrices divided the number of covariance matrices, and due to the specific property of this type of covariance matrices, the mean major eigenvector is obtained from the minor eigenvector of the mean covariance matrix. In summary, our proposed generalized multivariate test, which is quite similar to Eq.[1] in expression but different in interpretation, is expressed as follows:

$$(\mathbf{q}_1 - \overline{\mathbf{q}})^T \, \overline{\boldsymbol{\Sigma}}_{\overline{\mathbf{q}}}^{\pm} (\mathbf{q}_1 - \overline{\mathbf{q}}) \leq 2F(2, m; p) \,, \qquad\qquad [2]$$

where $\overline{\boldsymbol{\Sigma}}_{\overline{\mathbf{q}}}^{\pm}$ is the pseudoinverse of the mean covariance matrix, $\overline{\mathbf{q}}$ is the mean major eigenvector and $m$ is the average degrees of freedom, which is the arithmetic mean of the degrees of freedom associated with the collection of elliptical COUs. The key conceptual difference between Eq.[1] and Eq.[2] is the replacement of $\alpha$ with $p$. For a given test vector, $\mathbf{q}_1$, Eq.[2] can be used to determine the $p$-value, denoted by $p$, of the test.

By replacing the mean vector, $\overline{\mathbf{q}}$, and the pseudoinverse of the mean covariance matrix, $\overline{\boldsymbol{\Sigma}}_{\overline{\mathbf{q}}}^{\pm}$, of the control group with those of the single subject and the test vector, $\mathbf{q}_1$, with the mean vector of the control group, we would be able to compute the Type II error and the power of the proposed statistical test as well as the False Non-Discovery Rate (FNR) (Genovese and Wasserman, 2002), see Figure 2. For the sake of completeness, we rewrite Eq.[2] for this particular test as

$$(\overline{\mathbf{q}} - \mathbf{q}_1(\hat{\boldsymbol{\gamma}}))^T \boldsymbol{\Sigma}_{\mathbf{q}_1}^{+}(\hat{\boldsymbol{\gamma}})(\overline{\mathbf{q}} - \mathbf{q}_1(\hat{\boldsymbol{\gamma}})) \leq 2F(2, n-7; r) \,. \qquad\qquad [3]$$

Note that $r$ is used instead of $p$ is to avoid confusion but their meaning is similar. In short, Eq.[2] is a test of a single subject's eigenvector with respect to the mean eigenvector and the covariance matrix of the control group and Eq.[3] is a test of the control group's eigenvector with respect to the eigenvector and the covariance matrix of the single subject. Finally, while the framework is described within the context of single subject analysis, its generalization to group analysis should not pose any problem to interested readers.

*2.3 Novel Nonparametric Angular Dispersion (Shape) Deviation Indicator*

Shape characteristics of the elliptical COU we adopted here are the normalized areal measure and the normalized circumferential measure as proposed and derived in our earlier work (Koay et al., 2008). Differences between a single subject and a control group can then be tested as a one-against-many (or several-against-many) test based on the well-known nonparametric Wilcoxon-Mann-Whitney (WMW) test (Mann and Whitney, 1947; Wilcoxon, 1945), see Appendix B.

We would like to point out that a one-against-many test may require a lot more controls to reveal statistically significant outcomes than a several-against-many test. For example, there is only one single WMW U statistic, as defined in (Sachs, 1984), which is the first element with U=0, that has a probability less than 0.05 in a 1-against-45 test. It should be pointed out that we used the WMW U test as a two-tailed test. In the Results section, we used 4 independent samples as obtained from 4 separate sessions from each TBI patient to test them against a group of 45 controls. We also used 3 independent samples as obtained from 3 separate sessions from one non-TBI patient to test against the same control group. It also should be quite obvious that the WMW U statistic and its cumulative probability mass function can be used to estimate the number of controls needed perform a single-against-many or several-against-many test. A detail of the implementation of the computation of the exact *p*-value of the WMW U statistic  is further explored in the Discussion section and Figures 5 and 6.

## 3. RESULTS

We implemented the proposed framework in Java and performed several tests to evaluate the numerical accuracy of the proposed method and its implementation. The application of the inclusion-exclusion test as presented in Appendix C is useful for evaluating the accuracy of statistical models in representing the experimental data distributed on the sphere and for single-subject analyses. As an example, we utilized the inclusion-exclusion test to evaluate the fidelity of the $1-\alpha$ confidence region formed by an average (expected) elliptical COU of the major eigenvector of some specific tensor through Monte Carlo simulation. We also performed a simulation test to compare the relative merits of the two averaging methods, the arithmetic mean and the mean dyadics methods, for averaging a collection of covariance matrices of the major eigenvector. Finally, we illustrated the proposed methodology with clinical data.

*3.1 Testing Statistical Accuracy of the Elliptical COU*

The accuracy of the statistical framework depends directly on the validity of the elliptical COU as a measure of uncertainty of the major eigenvector. The inclusion-exclusion test developed in Appendix C provides the needed tool to quantify the fidelity of the elliptical COU as a model of dispersion for the major eigenvector. We carried out Monte Carlo simulation similar to that used in (Pierpaoli and Basser, 1996) to evaluate the fidelity of the elliptical COU. Specifically, we used a known tensor,

$$(D_{xx}, D_{yy}, D_{zz}, D_{xy}, D_{yz}, D_{xz}) = (9.475, 6.694, 4.829, 1.123, -0.507, -1.63) \times 10^{-4} \, mm^2/s \ ,$$

which has a fractional anisotropy (FA) of 0.4171, a known non-diffusion weighted signal of 1000 (arbitrary unit), and a DTI protocol based on the Sparse and Optimal Acquisition (SOA) 9x9 square design (Koay et al., 2012) (9 shells with 9 diffusion gradient directions on each shell) to generate noiseless diffusion-weighted signals and added Gaussian noise to these noiseless signals in quadrature (Henkelman, 1985; Koay and Basser, 2006) to obtain Rician-distributed noisy diffusion-weighted signals in this simulation study. The maximum b-value of 1500 s/mm$^2$ was applied to the outermost shell and the shells are equi-spaced in b-values. Rician-distributed noisy diffusion-weighted signals were generated from the diffusion tensor model with the above simulation parameters at a signal-to-noise ratio (SNR) of 20 (arbitrary unit). The SNR is defined conventionally as the non-diffusion weighted signal over the Gaussian noise standard deviation. Therefore, the *expected* variance, which is needed to constructed the average (or *expected*) 95% confidence region of the major eigenvector of the diffusion tensor is taken to be $\left(1000/SNR\right)^2$; Appendix IX of (Koay et al., 2007) outlines the construction of average covariance matrices for different quantities of interest.

The number of Monte Carlo trials was 20000. In each trial, a set of Rician-distributed diffusion-weighted signals was generated (Pierpaoli and Basser, 1996) and fitted with a diffusion tensor. We used the constrained nonlinear least squares estimation method for estimating the diffusion tensor (Koay et al., 2006). Figure 3 shows the estimated eigenvectors from these Monte Carlo trials as points on the unit sphere. The *expected* 95% confidence region of the major eigenvector is shown as a closed curve in white. The percent of eigenvectors within the expected confidence region as determined by the proposed inclusion-exclusion test was 95.08%, which was in very good agreement with the expected 95% used in this simulation study. The same simulation was repeated 500 times to quantify the confidence interval of the estimated percent of eigenvectors within the expected confidence region for different levels of SNR. The 99% confidence intervals of the percent of eigenvectors within the expected confidence region for SNRs of 15, 20, 25 and 30 were found to be (94.12%, 95.14%), (94.55%, 95.59%), (94.77%, 95.75%) and (94.88%, 95.84%), respectively.

*3.2 Testing Relative Merits of Methods for Averaging Covariance Matrices of the Major Eigenvector*

Adopting the same DTI protocol as in the previous simulation study, we investigated the relative merits of two different averaging approaches, the arithmetic mean and the mean dyadics methods, for computing the average covariance matrices of the major eigenvector of the diffusion tensor.  Since the number of Monte Carlo trials used the previous simulation study is equivalent to the number of samples used in computing the average covariance matrix of the major eigenvector, it should no longer be set at 20000 but at a number that is closer to the actual number of samples used in real experiments. Therefore, the number of samples (or Monte Carlo trials) in this simulation test was set at 45 because the sample size of our controls, which will be discussed next, was 45. The same simulation was repeated 500 times for different SNR levels to quantify the location and dispersion of the following 3 measures:

1. The Frobenius norm of the matrix difference between the averaged covariance matrix of the major eigenvector by means of either the arithmetic mean or mean dyadics methods and the expected covariance matrix of the major eigenvector as computed from the error propagation framework. Note that the lower the value of the Frobenious norm, the *closer* the averaged matrix to the expected matrix.

2. The relative error in estimating the areal and circumferential measures of the elliptical COU with respect to the ground truths, which were computed from the error propagation framework.

Table 1 shows the quantitative results of this simulation study. The numerical value quoted between the plus minus symbol is the standard deviation of the measurements. The results show clearly the arithmetic mean approach is uniformly better than the mean dyadics in computing the average covariance matrix of the major eigenvector.

*3.3 Illustration with Clinical Data*

We illustrate the application of the proposed framework by testing the data of two TBI patients and one single non-TBI subject against a control group of forty five non-TBI subjects. The TBI patients were volunteers in the National Capital Neuroimaging Consortium (NCNC) Neuroimaging Core project. Patient I and Patient II were 24 and 39 years old respectively at the time of their first scans. Subjects in this study were recruited and scanned between Aug. 2009 and May 2015. Non-TBI subjects (45 for the control group with mean age of $32.1\pm8.1$ years and 1 for the non-TBI single subject who was 22 years old) consisted of active-duty service members or dependents with no diagnosis of TBI and no history of other major neurologic disorders. The non-TBI subject population was a sample of convenience, comprised of any subject who met the eligibility requirements and without attempting to match for age, gender, education, etc. All scans were conducted with approval and according to the guidelines of the Walter Reed National Military Medical Center (WRNMMC) Institutional Review Board (IRB).

Each TBI patient was scanned in four separate occasions within a 3-year period. While some of the non-TBI subjects were scanned multiple times, we used only one session from each of the non-TBI subjects as our control group. Note that the single non-TBI subject used in this study is not part of the control group (N=45).

The diffusion-weighted images were collected as part of a larger protocol that included high-resolution anatomical images, functional MRI, spectroscopy, and Gd-perfusion. All MRI images were acquired on a 3T scanner (GE MR750, Milwaukee WI) with a 32-channel head coil. The diffusion-weighted images had the following parameters: TR≈10s, TE≈85ms, 2 $mm^3$ resolution, 49 diffusion-weighted gradient directions at b=1000 $s/mm^2$ and 6 b=0 $s/mm^2$ images, approximately 65 slices with data matrix of 128x128. A field map was collected, and cardiac gating was used to minimize cardiac motion artifacts.

The first phase of the DTIPrep (Oguz et al., 2014) quality control process was used to preprocess the imaging data. It was used to convert images from DICOM format to the NifTi format (http://nifti.nimh.nih.gov/), to check for slice-wise intensity artifacts, venetian blind noise, to average the non-diffusion weighted images and to remove diffusion measurements and the corresponding gradient directions that affected by artifacts. The data were then corrected for B0 distortion (Jezzard and Clare, 1999) and eddy currents with FSL (Jenkinson et al., 2012) using the measured field map. Data from gradient directions with excessive artifact as determined by DTIPrep were then omitted from the analysis. The summary statistics of the numbers of DWI measurements after DTIPrep is shown in Table 2. Two-sample tests of the numbers of DWI measurements after DTIPrep between the control group and TBI Patient I or the control group and TBI Patient II or the control group and the single non-TBI subject did not show any statistical significance, see the last column in Table 2. Therefore, it is unlikely the number of DWI measurements could be an nuisance effect. The Brain Extraction Tool (BET) in FSL (Jenkinson et al., 2012) was used prior to the estimation of the diffusion tensor. The diffusion tensor was estimated with the constrained nonlinear least squares method (Koay et al., 2006) as implemented in the HI-SPEED software packets (Koay, 2009. URL: http://sites.google.com/site/hispeedpackets/). The quality of the fit of the tensor to the data was evaluated using the reduced chi-square ($\chi^2$) statistic (Koay et al., 2006; Walker et al., 2011). The covariance matrix of the major eigenvector of the diffusion tensor was computed using our previously published analytical error propagation framework (Koay et al., 2007; Koay et al., 2008). Diffusion Tensor Imaging Toolkit (DTI-TK, http://dti-tk.sourceforge.net/) (Zhang et al., 2006), which is capable of preserving principal orientation of the diffusion tensor upon normalization, was used to normalize not only the diffusion tensors but also the covariance matrices of the major vector of the diffusion tensor of the control group and to construct an average tensor template with data matrix size of 256x256x128 at 1x1x1 mm$^3$ isotropic resampled resolution. The diffusion tensor data of TBI patients were then warped by DTI-TK to match the average template; the resultant displacement field map was then used to warp the covariance matrices of the major vector of the diffusion tensor of TBI patients to the space of the average tensor template. Finally, the elliptical cones of uncertainty of the major vectors were constructed from the warped covariance matrices of the major vectors of the diffusion tensors.

It should be noted that only those voxels from the average tensor template that had the following characteristics were considered for subsequent analysis:

(a) fractional anisotropy (FA) above 0.275,

(b) mean diffusivity above 2.5x10$^{-4}$ mm$^2$/s, and

(c) reduced $\chi^2$ statistic below or equal to a prescribed threshold. The $\chi^2$ cumulative distribution function (CDF) evaluated at this prescribed threshold is 0.95, see Appendix D for details.

The threshold value of FA at 0.275 was selected to ensure that the region of interest is mainly brain white matter. Furthermore, the voxel-wise averaged covariance matrix of the control group was set to null if more than 10 out of the 45 controls had a reduced $\chi^2$ statistic above the prescribed threshold. Otherwise, the voxel-wise averaged covariance matrix of the control group was computed from only those that had a reduced $\chi^2$ statistic below or equal to the prescribed threshold. The voxel-wise averaged covariance matrix of the individual subject was computed differently; it was set to null if at least one (out of four sessions) had a reduced $\chi^2$ statistic above the prescribed threshold.

*3.4 Novel FDR-FNR-based Orientation and Nonparametric Shape Deviation Results*

The FDR and FNR values were set at $1.0 \times 10^{-10}$ to ensure the *effect size* is large, i.e., the observed differences are likely to remain significant when the sample size increases. The results of the FDR-FNR-based Orientation Deviation Indicator based on the clinical data of two TBI patients and one non-TBI subject against the chosen group of 45 controls are shown in Figures 4A, 4B and 4C, respectively. The results shown in Figures 4A-C have been treated with cluster thresholding (Jenkinson et al., 2012) because of spatial smoothing, i.e,. image interpolation during image registration and tensor normalization. As such, the effective spatial resolution at which statistical inference can be made is generally larger than the acquired or resampled resolution. Therefore, only cluster of sufficient size (128 voxels) are shown in Figure 4. In the case of the Shape Deviation Indicator, the FDR was set at 0.0005 and the results for the two TBI patients are shown in Figures 4D and 4E. Similarly, cluster thresholding was applied to the results shown in Figures 4D-E. Interestingly, the Shape Deviation Indicator of the single non-TBI subject did not have any voxel that was statistically significant at this FDR threshold. Note that no statistically significant voxel was found before the application of cluster thresholding.

## 4. DISCUSSION

To the best of our knowledge, there are no clinical studies that uses the elliptical cones of uncertainty or its scalar measures (area and circumference of the elliptical cone) to characterize statistical features of fiber orientation. The goal of this work is to show how descriptive geometric measures could be utilized to detect changes in orientation or shape of the tracts as modeled by the elliptical cones of uncertainty in single-subject or group analyses. The salient features of the proposed framework are rigorous statistical quantification of orientation or shape deviation on a per-voxel basis, the incorporation of FDR and FNR methods for controlling the proportions of false positives and false negatives for an orientation deviation test and a nonparametric approach to testing shape deviation of the elliptical cones of uncertainty. Furthermore, a new implementation of computing the exact $p$-value of the WMW U statistic is presented.

Although the conceptual foundations of the proposed framework are simple and geometrically intuitive, the implementation is not trivial. It is made more complex by having to imbed it into a processing pipeline. The framework was developed and incorporated into our existing data analytics software tools, HI-SPEED software packets (Koay, 2009. URL: http://sites.google.com/site/hispeedpackets/). The whole processing pipeline was built on software tools not only from our group but also from other researchers who made their software tools publicly available, e.g., DTIPrep (Oguz et al., 2014) and DTI-TK (Zhang et al., 2006). In particular, the normalization of the covariance matrices of the major eigenvectors was made possible through a novel utilization of the tensor deformation tool within DTI-TK.

Ideally, any quantitative feature of interest that is to be used in a statistical test should be as statistically independent as possible from other features that might used by an image registration technique to drive the registration process because if this precautionary step can be carried out successfully it would remove image registration as a confound in any statistical analysis. However, it is almost an impossible feat in practice. Image registration is a real confound in practice and the problem of fully teasing apart this confound from any statistical analysis remains open and may not be tractable. Admittedly, the current framework is also not immune to the effects of this particular confound.

The most interesting preliminary biological finding from our clinical data is that the frontal portion of the superior longitudinal fasciculus seemed to be implicated in both tests (orientation and shape) as being significantly different from that of the control group. Another interesting result is that the Shape Deviation Indicator was able to separate the TBI patients from the single non-TBI subject at the chosen FDR level. The most puzzling result of this pilot study is that statistically significant voxels were found in the non-TBI subject under the orientation deviation test. Based on this preliminary test, we learned that the proposed orientation deviation test may be more sensitive to orientation changes in white matter tracts and perhaps at the cost of encountering more false positives. A follow-up study is planned to include more clinical data in single-subject analysis and to gain further insights from results across multiple single subjects.

An important aspect of this work was to study existing algorithms for computing the exact cumulative probability mass function of the Wilcoxon-Mann-Whitney U statistic. One of the current state-of-the-art numerical procedure is the method proposed by Nagarajan and Keich (Nagarajan and Keich, 2009). However, the method of Nagarajan and Keich is difficult to implement because of its dependence on a custom Fast Fourier transform method developed by the authors themselves. We discovered that the algorithm proposed by Harding (Harding, 1984), which is based on the method of generating functions and has the same complexity as that of Nagarajan and Keich, can be easily adapted with several minor modifications to computing the exact cumulative probability mass function of the Wilcoxon-Mann-Whitney U statistic in arbitrary-precision arithmetic. Interestingly, our implementation of the a slightly modified algorithm of Harding used only standard Java routines and was implemented with no more than several lines of codes, see Figure 5. Note that the algorithms shown in Figure 5 are based on exact executable Java codes. The Mathematica implementation of the same algorithm is even simpler than the Java implementation and is shown in Figure 6. Unfortunately, the method of generating functions cannot deal with tied ranks, i.e., data with ties. A different and less efficient approach has to be adopted, see Appendix F and online supplementary materials.

Finally, TOADDI is a new geometric and statistical framework for analyzing the elliptical COUs from DTI. TOADDI may be found useful in single-subject, graph-theoretic and group analyses of DTI data and the results from TOADDI may be included in DTI-based tractography techniques.

**Acknowledgment**

C.G. Koay dedicates this work to the memory of Madam Oh Soo See. The authors would like to thank Drs Connie Duncan and Louis French for sharing TBI patients' imaging data. The authors would also like to thank Ms. Elyssa Sham for coordinating the recruitment of patients and volunteers, Mr. John A. Morissette for acquiring the clinical data, Mr. Justin S. Senseney for managing the clinical data. Appendix E contains an abridged and concise version of the first part of the author's unpublished lecture note prepared for the occasion of the Mathematics/Computer Science Annual Alumni Lecture at Berea College on November 9, 2012. The current implementation of the proposed framework was built upon the HI-SPEED software packets of which C.G. Koay is the principal developer. HI-SPEED software packets are available for research use at URL: http://sites.google.com/site/hispeedpackets/. P.J. Basser and C. Pierpaoli are supported by the Intramural Research Program of the *Eunice Kennedy Shriver* National Institute of Child Health and Human  Development. The Mathematica® implementation of the two versions of the Wilcoxon-Mann-Whitney U tests, one with tied ranks and the other one without tied ranks, will be available as online supplementary materials on the publisher website.

**Appendix A.** *A step-by-step guide to constructing the covariance matrix of the major eigenvector of the diffusion tensor through constrained nonlinear least squares and to computing the normalized areal and circumferential measures of the elliptical cone of uncertainty*

This appendix outlines the most important steps from our previous works (Koay et al., 2006; Koay et al., 2007; Koay et al., 2008) that lead directly to the construction the covariance matrix of the major eigenvector of the diffusion tensor and its related scalar measures—the normalized areal and circumferential measures derived from the elliptical cone of uncertainty.

In DTI (Basser et al., 1994b), it is assumed that the diffusion-weighted signals can be modeled as follows:

$$s_i = S_0 \exp(-b_i \, \mathbf{g}_i^T \mathbf{D} \mathbf{g}_i), \qquad\qquad\qquad\qquad [A1]$$

where the measured signal, $s_i$, depends on some known quantities such as the diffusion encoding gradient vector, $\mathbf{g}_i$, which is of unit length and the diffusion weighting, $b_i$, and some unknown quantities such as the non-diffusion weighted signal, $S_0$, and the diffusion tensor, $\mathbf{D}$, which is a 3x3 symmetric positive definite matrix.

The nonlinear least squares method of estimation of the diffusion tensor can be formulated as a minimization problem with the following nonlinear least squares (NLS) objective function:

$$f_{NLS}(\boldsymbol{\gamma}) = \frac{1}{2}\sum_{i=1}^{m}\left( s_i - \exp(\textstyle\sum_{j=1}^{7} W_{ij}\,\gamma_j) \right)^2, \qquad\qquad\qquad [A2]$$

$$= \frac{1}{2}\sum_{i=1}^{m}\left( s_i - \hat{s}_i(\boldsymbol{\gamma}) \right)^2 = \frac{1}{2}\sum_{i=1}^{m} r_i^2(\boldsymbol{\gamma}) \quad, \qquad\qquad\qquad [A3]$$

where $m$ denotes the number of DWI measurements, $\gamma = [\ln(S_0), D_{xx}, D_{yy}, D_{zz}, D_{xy}, D_{yz}, D_{xz}]^T$ is the parameter vector to be estimated, $\hat{s}_i(\gamma) = \exp(\sum_{j=1}^{7} W_{ij}\gamma_j)$ is the model function evaluated at $\gamma$, $r_i = s_i - \hat{s}_i(\gamma)$ is one of the error terms of the NLS objective function, and the design matrix, $\mathbf{W}$, is given by

$$\mathbf{W} = \begin{pmatrix} 1 & -b_1 g_{1x}^2 & -b_1 g_{1y}^2 & -b_1 g_{1z}^2 & -2b_1 g_{1x}g_{1y} & -2b_1 g_{1y}g_{1z} & -2b_1 g_{1x}g_{1z} \\ \vdots & \vdots & \vdots & \vdots & \vdots & \vdots & \vdots \\ 1 & -b_m g_{mx}^2 & -b_m g_{my}^2 & -b_m g_{mz}^2 & -2b_m g_{mx}g_{my} & -2b_m g_{my}g_{mz} & -2b_m g_{mx}g_{mz} \end{pmatrix}.$$

It should be noted that the minimizer of the NLS objective function, which is denoted as $\hat{\gamma}$, may not necessarily produce a positive definite diffusion tensor estimate. To ensure that the diffusion tensor is at nonnegative definite, the subspace of the search space within the optimization process should not focus on the ordinary tensor elements but the Cholesky elements of the diffusion tensor (Koay et al., 2006; Pinheiro and Bates, 1996; Wang et al., 2004), i.e., $\mathbf{D} = \mathbf{U}^T\mathbf{U}$ and $\mathbf{U} = \begin{pmatrix} \rho_2 & \rho_5 & \rho_7 \\ 0 & \rho_3 & \rho_6 \\ 0 & 0 & \rho_4 \end{pmatrix}$. The parameter vector for the constrained nonlinear least squares (CNLS) objective function, $f_{CNLS}$, is $\boldsymbol{\rho} = [\rho_1, \rho_2, \rho_3, \rho_4, \rho_5, \rho_6, \rho_7]^T$ with $\rho_1 = \ln(S_0)$. Formally, $f_{CNLS}(\boldsymbol{\rho}) = f_{NLS}(\gamma(\boldsymbol{\rho}))$, $\gamma(\boldsymbol{\rho}) = [\rho_1, \rho_2^2, \rho_3^2 + \rho_5^2, \rho_4^2 + \rho_6^2 + \rho_7^2, \rho_2\rho_5, \rho_3\rho_6 + \rho_5\rho_7, \rho_2\rho_7]^T$ and the minimizer of the CNLS objective function is denoted by $\hat{\boldsymbol{\rho}}$. Although $\hat{\gamma} \neq \gamma(\hat{\boldsymbol{\rho}})$ when the diffusion tensor estimate is not positive definite, we shall assume for the sake of consistency with the main text that $\hat{\gamma}$ is obtained from $\gamma(\hat{\boldsymbol{\rho}})$ rather than obtained by minimizing the NLS objective function.

It can be shown that the covariance matrix of the major eigenvector of the diffusion tensor is given by

$$\mathbf{\Sigma}_{\mathbf{q}_1} = \mathbf{J}_{\boldsymbol{\gamma}}(\mathbf{q}_1)\mathbf{\Sigma}_{\boldsymbol{\gamma}}\mathbf{J}_{\boldsymbol{\gamma}}^T(\mathbf{q}_1),$$

where the eigenvalue (or spectral) decomposition of the diffusion tensor is given by

$$\mathbf{D} = \mathbf{Q}\mathbf{\Lambda}\mathbf{Q}^T = \sum_{i=1}^{3}\lambda_i \mathbf{q}_i \mathbf{q}_i^T \quad \text{with} \quad \lambda_1 \geq \lambda_2 \geq \lambda_3, \quad \text{the Jacobian matrix of } \mathbf{q}_1 \text{ with respect to the}$$

parameter vector, $\boldsymbol{\gamma}$, is given by

$$\mathbf{J}_{\boldsymbol{\gamma}}(\mathbf{q}_1) \equiv \mathbf{Q}\mathbf{T}_1,$$

$$\mathbf{T}_1 = \begin{pmatrix} 0 & 0 & 0 & 0 & 0 & 0 & 0 \\ 0 & \frac{1}{\lambda_1 - \lambda_2}\mathbf{a}(\mathbf{q}_2, \mathbf{q}_1)^T \\ 0 & \frac{1}{\lambda_1 - \lambda_3}\mathbf{a}(\mathbf{q}_3, \mathbf{q}_1)^T \end{pmatrix},$$

$$\mathbf{a}(\mathbf{q}_i, \mathbf{q}_j)^T \equiv [q_{ix}q_{jx}, q_{iy}q_{jy}, q_{iz}q_{jz}, q_{ix}q_{jy} + q_{iy}q_{jx}, q_{iy}q_{jz} + q_{iz}q_{jy}, q_{ix}q_{jz} + q_{iz}q_{jx}],$$

and the covariance matrix of the tensor elements can be expressed as a product of the estimate variance of the NLS fit and the inverse of the invariant Hessian matrix of the NLS function is given by,

$$\mathbf{\Sigma}_{\boldsymbol{\gamma}} = \sigma_{DW}^2 [inv(\nabla^2 f_{NLS}(\hat{\boldsymbol{\gamma}}))]^{-1},$$

$$= \sigma_{DW}^2 [\mathbf{W}^T(\hat{\mathbf{S}}^2 - \mathbf{R}\hat{\mathbf{S}})\mathbf{W}]^{-1}.$$

Note that $\sigma_{DW}^2 \equiv 2f_{NLS}(\hat{\boldsymbol{\gamma}})/(m-7)$. $\mathbf{R}$ and $\hat{\mathbf{S}}$ are diagonal matrices and their i-th diagonal elements are $r_i(\hat{\boldsymbol{\gamma}})$ and $\hat{s}_i(\hat{\boldsymbol{\gamma}})$, respectively. "$inv$" denotes the invariant part of the Hessian matrix, see (Koay et al., 2007).

Given the multivariate test as discussed in the main text, which is given by

$$(\mathbf{q}_1 - \mathbf{q}_1(\hat{\boldsymbol{\gamma}}))^T \boldsymbol{\Sigma}_{\mathbf{q}_1}^+(\hat{\boldsymbol{\gamma}})(\mathbf{q}_1 - \mathbf{q}_1(\hat{\boldsymbol{\gamma}})) \leq 2F(2, n-7; \alpha),$$

it can be shown (Koay et al., 2008) that its equivalent form can be expressed as

$\mathbf{q}_1^T \left(2F(2, n-7; \alpha)\boldsymbol{\Sigma}_{\mathbf{q}_1}(\hat{\boldsymbol{\gamma}})\right)^+ \mathbf{q}_1 \leq 1$. This inequality is geometrically an ellipsoidal region. Due to the unique construction of the covariance matrix of the major vector, this region is a degenerate ellipsoid because the length of its minor axis is zero. The minor axis is parallel to $\mathbf{q}_1(\hat{\boldsymbol{\gamma}})$ by design. Therefore, the spectral decomposition of $\boldsymbol{\Sigma}_{\mathbf{q}_1}(\hat{\boldsymbol{\gamma}})$ can be written as

$$\boldsymbol{\Sigma}_{\mathbf{q}_1}(\hat{\boldsymbol{\gamma}}) = \omega_1 \mathbf{c}_1 \mathbf{c}_1^T + \omega_2 \mathbf{c}_2 \mathbf{c}_2^T + 0\mathbf{q}_1(\hat{\boldsymbol{\gamma}})\mathbf{q}_1^T(\hat{\boldsymbol{\gamma}}).$$

The remaining two axes of the elliptical cone is of course orthogonal to $\mathbf{q}_1(\hat{\boldsymbol{\gamma}})$ and their lengths are $a \equiv \sqrt{2F(2, n-7; \alpha)\omega_1}$ and $b \equiv \sqrt{2F(2, n-7; \alpha)\omega_2}$, respectively. Therefore, the degenerate ellipsoid is reduced to an ellipse on the plane orthogonal to $\mathbf{q}_1(\hat{\boldsymbol{\gamma}})$, which is similar to that shown in Figure 1A in which $\mathbf{q}_1(\hat{\boldsymbol{\gamma}})$ is pointing along positive z axis.

The normalized areal and circumferential measures can be computed from the following expressions, which depend only on $a$ and $b$,

$$\Gamma(a,b) = \frac{2a}{\pi b \sqrt{1+a^2}} \left((1+b^2)\Pi(-b^2, \beta) - \mathrm{K}(\beta)\right), \text{ and}$$

$$\Lambda(a,b) = \frac{2}{\pi b \sqrt{1+a^2}} \left((1+b^2)\Pi(\beta, \omega) - \mathrm{K}(\omega)\right), \text{ respectively,}$$

where $\omega = \frac{b^2 - a^2}{b^2(1+a^2)}$, $\beta = \frac{a^2 - b^2}{1+a^2}$. Further, $\mathrm{K}$ and $\Pi$ are the complete elliptic integrals of the first kind and the third kind, respectively, and their definitions are stated in (Koay et al., 2008). Please refer to online supplementary materials for a numerical example.

**Appendix B.** *The Wilcoxon-Mann-Whitney U Statistic*

In this appendix, we give a brief review of the computation of the Wilcoxon-Mann-Whitney U statistic from two independent samples, $\mathbf{x}$ and $\mathbf{y}$, and discuss the concept of the null probability mass distribution function of the U statistic from which the p-value of the Wilcoxon-Mann-Whitney U statistic can computed. The null probability mass distribution function is derived from the assumption that elements of $\mathbf{x}$ and $\mathbf{y}$ are equally likely to appear in any order.

Let $\mathbf{x}$ be an array with $m$ elements and $\mathbf{y}$ be another array with $n$ elements. We assume that every element in $\mathbf{x}$ or $\mathbf{y}$ can ordered by the order relation $<$. Under this order relation, it is assumed that the probability is zero of finding any two elements to be of the same numerical value. In other words, we will not discuss U statistics with tied ranks, see Appendix F on this topic.

Combine $\mathbf{x}$ and $\mathbf{y}$ into a single array and sort this combined array in an ascending order while keeping track of each element's original membership to either $\mathbf{x}$ or $\mathbf{y}$. Assign indices from $1$ to $m+n$ to this sorted combined array in ascending order. Sum the indices of those elements originally belonged to $\mathbf{x}$ and let the sum be $R_1$. Similarly, sum the indices of those elements originally belonged to $\mathbf{y}$ and let the sum be $R_2$. The test statistic $U$ is given by:

$$U = \min(U_1, U_2),$$ [B1]

where

$$U_1 = m\,n + \frac{m(m+1)}{2} - R_1,$$ [B2]

and

$$U_2 = m\,n + \frac{n(n+1)}{2} - R_2 \ \text{ or}$$ [B3]

$$U_2 = m\,n - U_1.$$ [B4]

The above definition of the test statistic is consistent with the one presented in (Sachs, 1984).

Under the assumption that any element of $\mathbf{x}$ or $\mathbf{y}$ is equally likely to appear in anywhere in a combined array of length $m+n$, the probability mass distribution can be found in the following straightforward procedure. This counting strategy is conceptually simple but is not recommend as a computational tool in practice except for small sample sizes, e.g., $m+n \leq 12$. Another slightly more efficient method of counting is described in Appendix F to deal measurements with ties.

First, assign $1$ to $m+n$ to an array. Second, generate all the possible permutations from this array. Third, we perform the following steps for each permutation; assign the first $m$ elements to $\mathbf{x}$ and the remaining $n$ elements to $\mathbf{y}$, and compute the test statistic $U$. Finally, the probability mass distribution function of the Wilcoxon-Mann-Whitney statistic at any value of $U$ is simply the ratio of the number of times that specific value test statistic appears among all the possible permutations to the total number of permutations. It should be clear that the cumulative probability mass distribution function can be obtained from its probability mass distribution function.

The above counting strategy may be used to verify the correctness of one's algorithm that is based Algorithm C of Figure 5 for small sample sizes. Efficient computation of the cumulative probability mass distribution function of the Wilcoxon-Mann-Whitney statistic should of course be based on Algorithm C as shown in Figure 5 or the Mathematica implementation of the same algorithm as shown in Figure 6. A thorough exposition on the application of generating function and the theory of partitions of positive integers to the computation of the cumulative probability mass distribution function of the Wilcoxon-Mann-Whitney statistic is interesting but it is beyond the scope of this paper.

**Appendix C.** *Inclusion-exclusion Test: Numerical Implementation*

To test whether a point (represented as a unit vector) on the unit sphere lies inside an elliptical COU of interest, it is most efficient to reorient the elliptical COU to its canonical position in which the center of the cone is aligned parallel with respect to the positive z-axis and the major and minor axes of the ellipse of the cone are aligned with the positive x and positive y axes, respectively, see Figure 1A and 1B. This reorientation is formed simply as a product of two rotation matrices. Once the necessary rotation matrix is found, it can be applied to the point of interest. The rotated point (or vector) of interest is then projected onto the u-v plane and the inclusion-exclusion test is reduced to testing whether the point is inside the ellipse of the u-v plane or otherwise. It should be clear that such an inclusion-exclusion test can be determined very efficiently because it is simply an inequality test, which will be outlined here.

Let $a\mathbf{c}_1$, $b\mathbf{c}_2$ and $\mathbf{q}$ be the vectorial representation of the major axis, the minor axis and the center of the ellipse of an elliptical COU. Further, let $\hat{\mathbf{x}}, \hat{\mathbf{y}}$ and $\hat{\mathbf{z}}$ be the unit vectors along the positive x, y and z axes, respectively. Rotating $\mathbf{q}$ to be along $\hat{\mathbf{z}}$ can be accomplished by the following rotation matrix,

$$\mathbf{M}_1 = \Re(\cos^{-1}(\mathbf{q}.\hat{\mathbf{z}}), \frac{\mathbf{q} \times \hat{\mathbf{z}}}{\|\mathbf{q} \times \hat{\mathbf{z}}\|}) \ . \tag{C1}$$

Please refer to Appendix E for further details on the rotation representation used in this work. The rotated vector of the inverse Gnomonic projection of $\mathbf{c}_1 + \mathbf{q}$ is simply the normalized vector,

$$\mathbf{r}_1 = \mathbf{M}_1 \frac{\mathbf{c}_1 + \mathbf{q}}{\|\mathbf{c}_1 + \mathbf{q}\|} . \tag{C2}$$

Aligning $\mathbf{r}_1$ in such a way that it lies on the x-z plane and with its orientation pointing in the positive x axis can be accomplished by another rotation, $\mathbf{M}_2$, about $\hat{\mathbf{z}}$, which is given by

$$\mathbf{M}_2 = \Re(\phi, \hat{\mathbf{z}}) , \tag{C3}$$

where $\phi$ is the second component of the spherical coordinates, $(\theta, \phi)$, of $\mathbf{r}_1$. Note that the transformation between the Cartesian coordinates and the spherical coordinates follows the convention commonly used in Physics: $x = r\sin(\theta)\cos(\phi)$, $y = r\sin(\theta)\sin(\phi)$, and $z = r\cos(\theta)$.

Once the above two rotation matrices are found, the inclusion-exclusion test of a point on the unit sphere can be determined efficiently by rotating the point to the canonical position of the elliptical COU and then project the rotated point onto the canonical u-v plane of the elliptical COU. For example, let $\mathbf{p}$ be a point on the unit sphere. To test whether it is inside the elliptical COU of interest, we rotate $\mathbf{p}$ to the canonical position of the elliptical COU by

$$\tilde{\mathbf{p}} = \mathbf{M}_2 \mathbf{M}_1 \mathbf{p} . \qquad\qquad [C4]$$

Projecting $\tilde{\mathbf{p}}$ onto the canonical u-v plane can be accomplished by the Gnomonic projection:

$$\mathbf{p}' = \tilde{\mathbf{p}} / \tilde{\mathbf{p}} \cdot \hat{\mathbf{z}} . \qquad\qquad [C5]$$

The first step in the inclusion-exclusion test is to test the following condition, $\tilde{\mathbf{p}} \cdot \hat{\mathbf{z}} \leq 0$. If the condition is true, $\mathbf{p}$ is not inside the elliptical COU. Otherwise, we need to test the following condition:

$$\left(\frac{p'_x}{a}\right)^2 + \left(\frac{p'_y}{b}\right)^2 \leq 1 , \qquad\qquad [C6]$$

where $\mathbf{p}' = [p'_x, p'_y, p'_z]$.

It should be clear that once the inclusion-exclusion test can carried out for a single elliptical COU, see Figure 2 for an example, the inclusion-exclusion test for a union of a collection of elliptical COUs can be performed similarly in the non-ideal case that a collection of elliptical COUs may not be drawn from the same distribution. In practice, it is not necessary to rotate the elliptical cone; it is only necessary to extract the rotation matrix, $\mathbf{M}_2 \mathbf{M}_1$, and apply it to the point (or vector) of interest. One fact that should be incorporated into the implementation of the inclusion-exclusion test to enhance efficiency is that a point is outside the elliptical COU if it is not within the outer circular cone formed by the major axis of the elliptical COU.

The combined rotation matrix, $\mathbf{M}_2\mathbf{M}_1$ , maps an elliptical COU to its canonical position. One important application of the inverse of this combined rotation matrix, $\mathbf{M}_1^T\mathbf{M}_2^T$ , is in visualization. Specifically, it is more convenient to generate polygonal (graphical) representations of these elliptical COUs in their canonical positions and then rotate these representations to their original positions. Figure 3 shows an example in which the confidence region (in white) was transformed by the inverse of this combined rotation matrix to its original position to show how well this 95% confidence region captured the dispersion of the estimated eigenvectors.

**Appendix D.** *Reduced $\chi^2$ statistic*

The reduced $\chi^2$ statistic is the estimated noise variance from the nonlinear least square fit divided by a constant (Koay et al., 2006). This constant is typically chosen to be the global noise variance, which is a known quantity in simulation studies (Koay et al., 2006) or an estimated quantity in experimental studies (Walker et al., 2011).

We express the $\chi^2$ probability density function with $\nu$ degrees of freedom as follows:

$$g_{\chi^2}(x)\,dx = \frac{x^{\nu/2-1}}{\Gamma(\nu/2)\,2^{\nu/2}}\exp(-x/2)\,dx, \quad 0 < x < \infty, \qquad [\text{D1}]$$

where $\Gamma$ is the Euler Gamma function. In the context of this work, the degrees of freedom is the difference between the number of diffusion measurements used in the fitting and the number of estimated parameters, which is 7 in DTI. The reduced $\chi^2$ probability density function with $\nu$ degrees of freedom is given by:

$$g_{\chi^2_\nu}(x)\,dx = \nu\, g_{\chi^2}(\nu x)\,dx\,. \qquad [\text{D2}]$$

We would like to derive the critical threshold based on a prescribed right-tail area of the reduced $\chi^2$ distribution. Let the prescribed right-tail area be $\alpha$ and the critical threshold be $\beta$, the relationship between these two quantities is given by

$$\alpha = \int_\beta^\infty \nu\, g_{\chi^2}(\nu x)\,dx\,,$$

$$= \int_{\beta\nu/2}^\infty \frac{t^{\nu/2-1}}{\Gamma(\nu/2)}e^{-t}\,dt\,,$$

$$= Q(\nu/2, \beta\nu/2)\,,$$

where $Q(a,x) = \frac{\Gamma(a,x)}{\Gamma(a)}$ and $\Gamma(a,x) \equiv \int_x^\infty t^{a-1} e^{-t} \, \mathrm{d}t$ is the upper incomplete Gamma function.

The change of variables from $x$ and $t$ above was based on this equation, $t = \nu\, x / 2$. Finally, the critical threshold, $\beta$, can be expressed as

$$\beta = \frac{2}{\nu} Q^{-1}(\nu/2, \alpha),$$

where $Q^{-1}(a,s)$ is the solution $x$ to $Q(a,x) = s$.

**Appendix E.** *Matrix representation of a rotation about an axis*

The matrix representation of a rotation about an axis is due to L. Euler (Euler, 1776). Let an axis be represented by a unit vector $\hat{\mathbf{n}} = [n_1 \quad n_2 \quad n_3]^T$, i.e., $n_1^2 + n_2^2 + n_3^2 = 1$. Suppose that the unit vector, $\hat{\mathbf{n}}$, is pointing out of the page. The counter-clockwise rotation (as viewed by the reader looking at the page) about $\hat{\mathbf{n}}$ of angle $\Phi$, which we denote as $\Re(\Phi, \hat{\mathbf{n}})$, is given by:

$$\Re(\Phi, \mathbf{n}) = \begin{pmatrix} t\,n_1^2 + c & t\,n_1 n_2 - s\,n_3 & t\,n_1 n_3 + s\,n_2 \\ t\,n_1 n_2 + s\,n_3 & t\,n_2^2 + c & t\,n_2 n_3 - s\,n_1 \\ t\,n_1 n_3 - s\,n_2 & t\,n_2 n_3 + s\,n_1 & t\,n_3^2 + c \end{pmatrix},$$ [E1]

where $t = 1 - c$, $c = \cos(\Phi)$, and $s = \sin(\Phi)$.

For completeness, we outline a physically motivated derivation of Eq.[E1] through the equation of the counter-clockwise constant precession (with angular velocity, $\mathbf{\Omega}$) of a time-dependent vector, $\mathbf{r}(t)$, around an axis, $\hat{\mathbf{n}}$, i.e., $\hat{\mathbf{\Omega}} = \hat{\mathbf{n}}$. Let $\omega \equiv \|\mathbf{\Omega}\| \equiv \frac{d\phi}{dt}$, constant precession in this context is given by $\frac{d\mathbf{\Omega}}{dt} = \mathbf{0}$, which implies that (1) $\frac{d\omega}{dt} = 0$ and $\frac{d\hat{\mathbf{\Omega}}}{dt} = \mathbf{0}$. The equation of precession is usually given in vector notation,

$$\frac{d}{dt}\mathbf{r}(t) = \mathbf{\Omega} \times \mathbf{r}(t).$$ [E2]

Equation [E2] can be expressed in matrix notation as a linear homogeneous system of differential equations with constant coefficients as

$$\frac{d}{dt}\begin{pmatrix} r_1(t) \\ r_2(t) \\ r_3(t) \end{pmatrix} = \omega \underbrace{\begin{pmatrix} 0 & -n_3 & n_2 \\ n_3 & 0 & -n_1 \\ -n_2 & n_1 & 0 \end{pmatrix}}_{\mathbf{N}} \begin{pmatrix} r_1(t) \\ r_2(t) \\ r_3(t) \end{pmatrix}.$$ [E3]

Suppose that during the period from $0$ to $\tau$, the vector, $\mathbf{r}(t)$, moves from $\mathbf{r}(0)$ to $\mathbf{r}(\tau)$. Since $\omega = \frac{d\phi}{dt}$ and $\omega$ is a constant, the angle swept out by the precession during this period is $\Phi \equiv \int_0^\Phi d\phi = \int_0^\tau \omega dt = \omega\tau$. The solution to Eq.[E3] can be solved formally and expressed in matrix exponential form (Hirsch and Smale, 1974), which is given by ,

$$\begin{pmatrix} r_1(t) \\ r_2(t) \\ r_3(t) \end{pmatrix} = \exp(\Phi\,\mathbf{N}) \begin{pmatrix} r_1(0) \\ r_2(0) \\ r_3(0) \end{pmatrix}.$$

By expanding the matrix exponential in series representation, it can be shown that

$$\exp(\Phi\,\mathbf{N}) = \mathbf{I} + \sin(\Phi)\mathbf{N} + (1 - \cos(\Phi))\mathbf{N}^2 , \qquad\qquad \text{[E4]}$$

where $\mathbf{I}$ is the 3x3 identity matrix. $\exp(\Phi\,\mathbf{N})$ is a proper rotation matrix, i.e., $\exp(\Phi\,\mathbf{N})\left[\exp(\Phi\,\mathbf{N})\right]^T = \exp(\Phi\,\mathbf{N})\exp(\Phi\,\mathbf{N}^T) = \exp(\Phi\,\mathbf{N})\exp(-\Phi\,\mathbf{N}) = \mathbf{I}$ and $\det(\exp(\Phi\,\mathbf{N})) = \exp(\Phi\,tr(\mathbf{N})) = \exp(0) = +1$, see page 171 of (Bellman, 1970) for the derivation of the Jacobi identity, i.e., $\det(\exp(\mathbf{M})) = \exp(tr(\mathbf{M}))$ for some square matrix $\mathbf{M}$. Note that $\det(\cdot)$ and $tr(\cdot)$ denote the matrix determinant and the matrix trace, respectively. Finally, Eq.[E1] follows directly from Eq.[E4].

**Appendix F.** *Treatment of ties in the Wilcoxon-Mann-Whitney test*

The treatment of ties in computing the Wilcoxon-Mann-Whitney U statistic can be accomplished by assigning the average of those ranks that share the same numerical value, e.g., see (Lehmann and D'Abrera, 2006). For example, if the second and the third elements in the combined ordered array are the same in numerical value their ranks should be assigned the average value of their of ranks, which is (2+3)/2=2.5. Once tied ranks have been assigned, the Wilcoxon-Mann-Whitney U statistic can be computed without any modification as described in Appendix B. Since the null probability mass distribution function of the Wilcoxon-Mann-Whitney U statistic with tied ranks is different from that of the Wilcoxon-Mann-Whitney U statistic without tied ranks, it has to be computed afresh. Herein, we describe another counting strategy for the purposes of computing the null probability mass distribution function of the Wilcoxon-Mann-Whitney U statistic with tied ranks.

Let $\mathbf{x}$ be an array with $m$ elements and $\mathbf{y}$ be another array with $n$ elements. Without loss of generality, let's assume that $m \leq n$. Combine $\mathbf{x}$ and $\mathbf{y}$ into a single array with $m+n$ elements and compute the ranks of this combined array. Given $m+n$ tied ranks, we can generate all the combinations with $m$ elements chosen from $m+n$ tied ranks. The total number of combinations is the binomial coefficient, $C_m^{m+n}$. For each combination of $m$ elements, compute the Wilcoxon-Mann-Whitney U statistic. The probability mass distribution function of the Wilcoxon-Mann-Whitney statistic at any value of $U$ is simply the ratio of the number of times that specific value test statistic appears among all the possible combinations to the total number of combinations.

Table and Figure Captions

Table 1. Quantitative comparison of two different averaging methods.

| SNR | 15 | | 20 | | 25 | | 30 | |
|---|---|---|---|---|---|---|---|---|
| Methods<br>Measures | Arith.<br>Mean | Mean<br>Dyadics | Arith.<br>Mean | Mean<br>Dyadics | Arith.<br>Mean | Mean<br>Dyadics | Arith.<br>Mean | Mean<br>Dyadics |
| Frobenius<br>Norm | $7.3 \times 10^{-4}$<br>$\pm 4.4 \times 10^{-4}$ | 0.001<br>$\pm 0.002$ | $2.5 \times 10^{-4}$<br>$\pm 1.5 \times 10^{-4}$ | 0.001<br>$\pm 1.5 \times 10^{-4}$ | $1.3 \times 10^{-4}$<br>$\pm 7.3 \times 10^{-5}$ | $6.8 \times 10^{-4}$<br>$\pm 3.5 \times 10^{-4}$ | $7.5 \times 10^{-5}$<br>$\pm 4.4 \times 10^{-5}$ | $4.6 \times 10^{-4}$<br>$\pm 2.2 \times 10^{-4}$ |
| Relative Error<br>in AM | 0.060<br>$\pm 0.042$ | 0.123<br>$\pm 0.091$ | 0.038<br>$\pm 0.028$ | 0.121<br>$\pm 0.088$ | 0.031<br>$\pm 0.023$ | 0.111<br>$\pm 0.088$ | 0.028<br>$\pm 0.020$ | 0.121<br>$\pm 0.087$ |
| Relative Error<br>in CM | 0.032<br>$\pm 0.022$ | 0.067<br>$\pm 0.049$ | 0.020<br>$\pm 0.015$ | 0.062<br>$\pm 0.045$ | 0.017<br>$\pm 0.012$ | 0.060<br>$\pm 0.045$ | 0.014<br>$\pm 0.010$ | 0.062<br>$\pm 0.045$ |

AM = Areal Measure, CM = Circumferential Measure, SNR = Signal-to-Noise Ratio

Table 2. Summary statistic of the number of DWI measurements after the DWI images were processed by DTIPrep. The data in the last column is the p-values of the two-sample test based on the Wilcoxon-Mann-Whitney method of the numbers of DWI measurements between the control group and TBI Patient I, TBI Patient II and Non-TBI subject, respectively. Note that the computation of the exact p-value of the WMW U statistic with tied ranks was based on the method described in Appendix F.

| summary statistic / subject | Max. | Min. | Mean ± SD | Two-sample p-value via WMW |
|---|---|---|---|---|
| Controls | 49 | 43 | 47.73 ± 1.48 | --- |
| TBI Patient I | 49 | 44 | 47.00 ± 2.16 | U=70, p=0.4822 |
| TBI Patient II | 49 | 42 | 46.75 ± 3.20 | U=77, p=0.5998 |
| Non-TBI subject | 49 | 48 | 48.33 ± 0.58 | U=58, p=0.7628 |

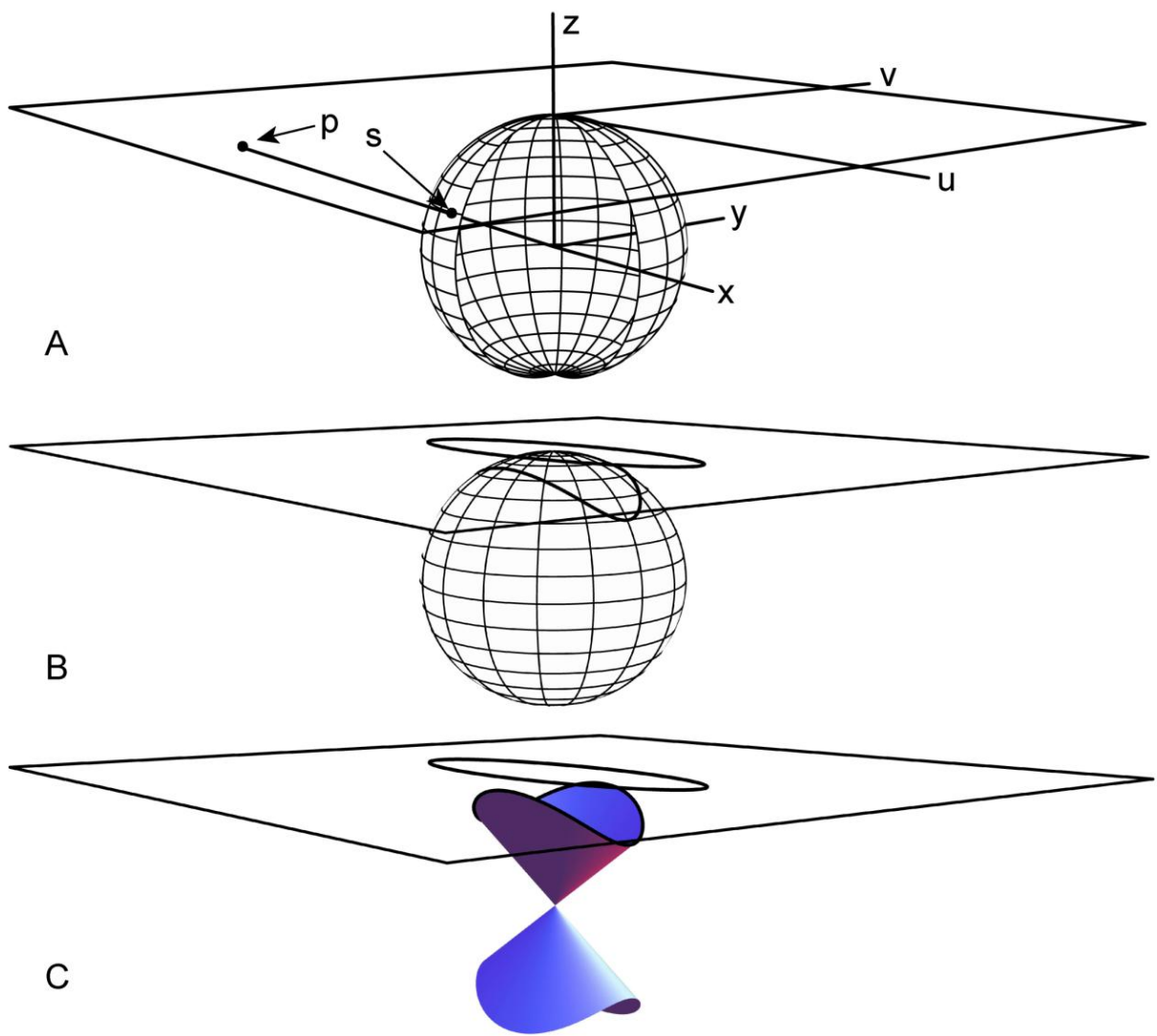

Figure 1. (A) The Gnomonic projection maps a point, s, on the unit sphere to a point, p, on the u-v plane in which the line connecting both points also passes through the center of the unit sphere. The u-v plane is tangent to the unit sphere at (0,0,1). The inverse Gnomonic projection is the inverse of the Gnomonic projection. (B) An ellipse on the u-v plane and its inverse Gnomonic projection on the unit sphere. (C) The elliptical cone of uncertainty with its center aligned along the z-axis.

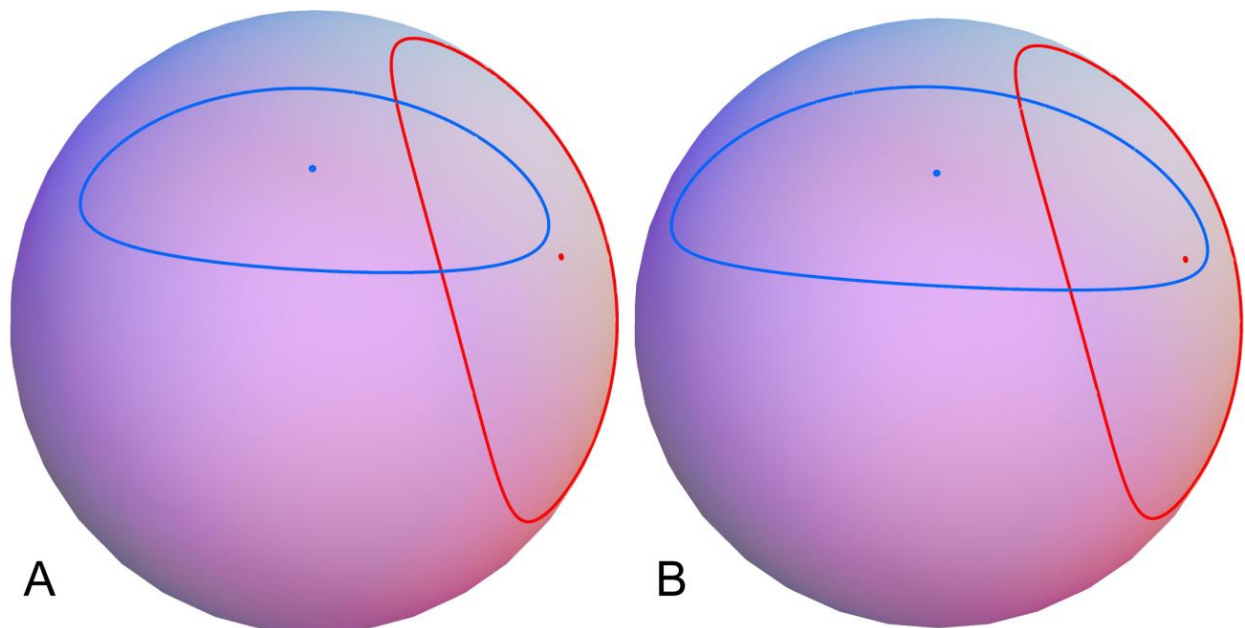

Figure 2. Given two distinct centers of the elliptical COUs, one of the control group (red) and the other of an individual subject (blue), two scenarios may occur if the size or the orientation of the elliptical COU of the individual subject varies differently. In both cases (A and B), the center of an individual subject is deemed to be significantly different in a statistical sense from the center of the control group. However, the first scenario (A) has a lower Type II error than the second scenario (B) because the center of the control group is also deemed significantly different from that of the individual subject. In (B), the center of the control group is not significantly different from that of the individual subject in the second scenario because it is located within the confidence cone of the individual subject.

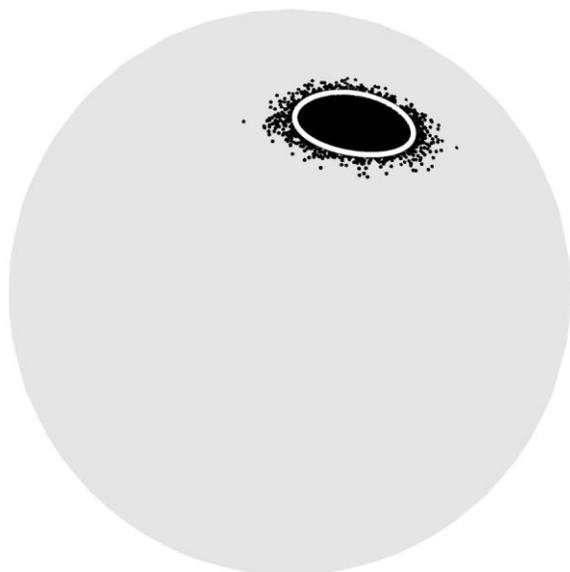

Figure 3. A collection of 20000 estimated eigenvectors produced from the constrained nonlinear least squares estimation of the diffusion tensor. The expected 95% confidence region from our previous proposed error propagation framework is shown in white. The percent of eigenvectors within the expected confidence region was 95.08% as determined by the inclusion-exclusion test.

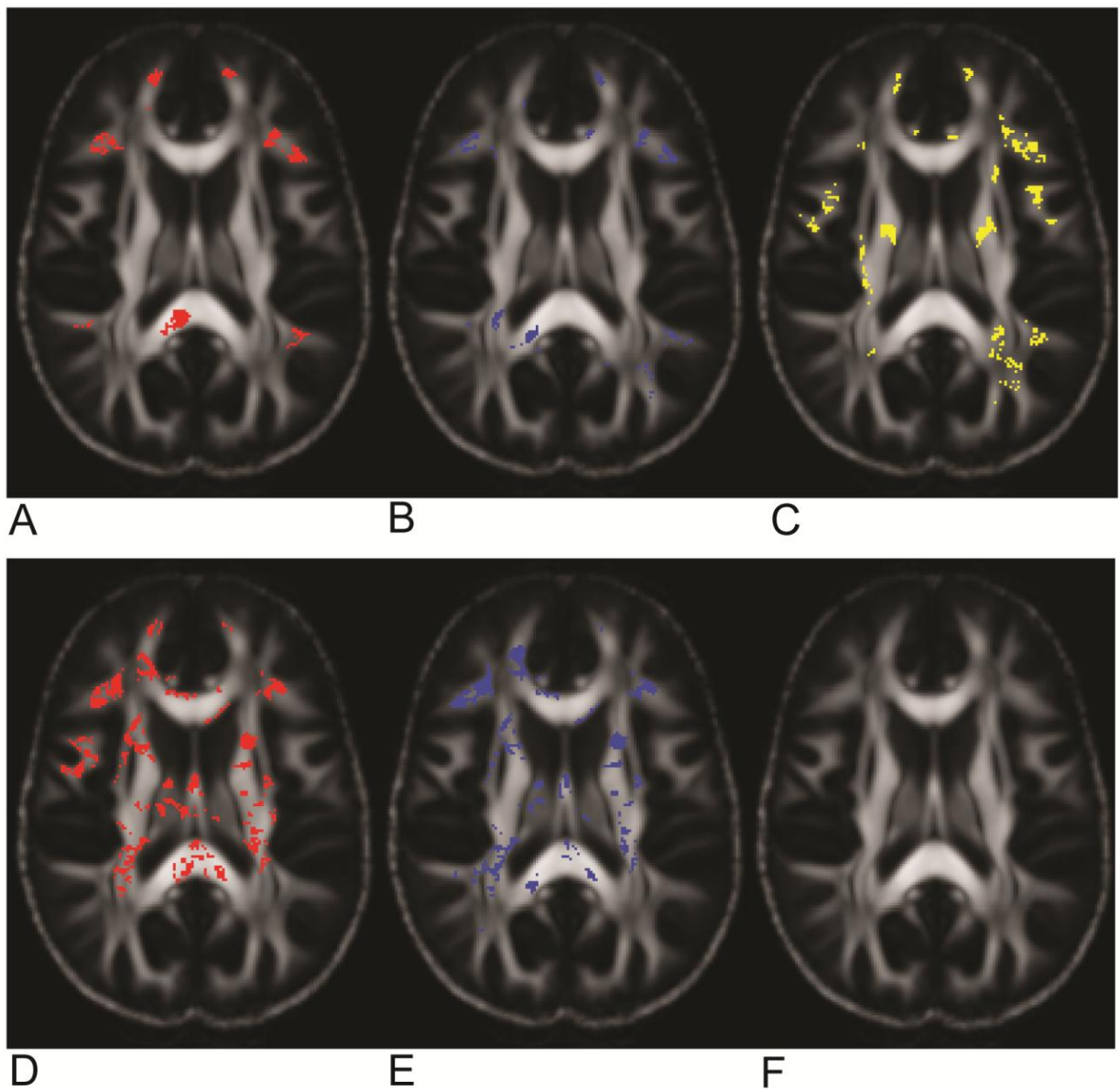

Figure 4. Results of statistically significant deviation in orientation shown in (A) red, (B) blue and (C) yellow for TBI patient I, TBI patient II and single non-TBI subject, respectively. Results of statistically significant deviation in shape of the elliptical COU shown in (D) red and (E) blue for TBI patient I and TBI patient II, respectively. No voxel was found to be statistically significant under the shape deviation test for the non-TBI subject (F).

```
public static BigInteger factorial(int aa) {

    if (aa==0 || aa==1) return BigInteger.ONE;

    BigInteger a = BigInteger.valueOf(aa);
    BigInteger r = BigInteger.ONE;
    BigInteger term = BigInteger.ONE.add(BigInteger.ONE);

    while(term.compareTo(a)<=0){
        r = r.multiply(term);
        term = term.add(BigInteger.ONE);
    }

    return r;
}
```

Algorithm A: Factorial function

```
public static BigInteger binomial(int n, int r){

    if (r>n) return BigInteger.ZERO;
    else if (r > (n-r)) return binomial(n,n-r);
    else if ( n>=0 && r==0 ) return BigInteger.ONE;
    else if ( (n>=0 && r==(n-1)) || (n>=0 && r==1) )
        return BigInteger.valueOf(n);
    else if (n==r) return BigInteger.ONE;

    BigInteger numerator = BigInteger.valueOf(n);
    BigInteger term = BigInteger.valueOf(n-1);

    int rr = r;
    while(rr>=2){
        numerator = numerator.multiply(term);
        term = term.subtract(BigInteger.ONE);
        rr=rr-1;
    }

    return numerator.divide(factorial(r));
}
```

Algorithm B: Binomial function

```
public static BigDecimal[] GetPValues(int nn, int mm){

    int m = Math.max(nn,mm);
    int n = Math.min(nn,mm);

    int M =  ((m*n)%2==0)? (m*n)/2 : (m*n+1)/2 -1 ;

    BigInteger[] f = new BigInteger[M+1];
    for(int i=0; i<M+1; i++) f[i] = BigInteger.ONE;

    BigInteger b = binomial(n+m, n);

    int Q = Math.min(m, M);
    for(int s = 1; s<=Q; s++)
        for(int u = s; u<=M; u++)
            f[u] = f[u].add( f[u-s]);

    int P = Math.min(n+m, M);
    for(int t = n+1; t<=P; t++)
        for(int u = M; u>=t; u--)
            f[u] = f[u].subtract( f[u-t]);

    BigDecimal[] cpmf = new BigDecimal[M+1];

    for(int i=0; i<M; i++) {
        cpmf[i] = new BigDecimal(f[i].add(f[i]));
        cpmf[i] = cpmf[i].divide(new BigDecimal(b),MathContext.DECIMAL128);
    }
    cpmf[M] = BigDecimal.ONE;

    return cpmf;
}
```

Algorithm C: The cumulative probability mass distribution
function of the null distribution of the
two-sample Wilcoxson-Mann-Whitney U statistic.

Figure 5: The algorithm for computing exact *p*-value of the two-sample Wilcoxon-Mann-Whitney U statistic in arbitrary and high precision is shown in Algorithm C. This algorithm depends on the Binomial function and the Binomial function in turns depends on the Factorial function. Both of the auxiliary functions are listed as Algorithms A and B. The above algorithms are based on exact Java codes and can be executed once they are placed within a Java class. By definition, the U statistic is derived from a two-tailed test. It should be clear that the *p*-value can also be represented in rational representation by a slight modification of Algorithm C once a class for representing the rational numbers has been created. It should be noted that the Java static method CPMF returns an array of *p*-values. Note that the array indexing in Java begins at 0. For example, the probability that the U statistic is less than or equal to $U_c$ is simply the $U_c^{th}$ element of the returned array. The last element of the array is always unity.

```
GetPValues[m_, n_] :=
Module[{mm, nn, M, f, d, p, q, s, u, t, i},

If[m >= n,
      nn = m;  mm = n,
      nn = n; mm = m;
  ];

If[ Mod[mm*nn, 2] == 0,
     M = (mm*nn)/2;,
     M = (mm*nn + 1)/2 - 1;
  ];

f = Table[1, {M + 1}];
d = Binomial[nn + mm, nn];

q = Min[mm, M];
For[s = 1, s <= q, s++,
 For[u = s, u <= M, u++,
   f[[u + 1]] = f[[u + 1]] + f[[u - s + 1]];
   ];
  ];

p = Min[mm + nn, M];
For[t = nn + 1, t <= p, t++,
 For[u = M, u >= t, u--,
   f[[u + 1]] = f[[u + 1]] - f[[u - t + 1]];
   ];
  ];

For[i = 1, i <= M, i++,
 f[[i]] = 2*f[[i]]/d;
  ];
f[[M + 1]] = 1;

Transpose[{Range[M + 1] - 1, N[f, 20]}]
]
```

Figure 6: The Mathematica® implementation of the same algorithm for computing exact *p*-value of the two-sample Wilcoxon-Mann-Whitney U statistic as presented in Algorithm C of Figure 5. If higher precision is desired, the number of significant digits can be increased by changing "20" to a larger integer on the last statement, e.g., replace "N[f,20]" with "N[f,30]" if 30 significant digits is desired.